\documentclass[%
 reprint,
superscriptaddress,
 amsmath,amssymb,
 aps,
floatfix,
]{revtex4-2}

\usepackage{latexsym}
\usepackage{graphicx}
\usepackage{dcolumn}
\usepackage{bm}
\usepackage{xcolor}

\begin{document}


\title{High-Field Brightness Limits of Alkali-Antimonide Photocathodes}

\author{Peng-Wei Huang}
\altaffiliation{Email: hpw17@tsinghua.org.cn}
\affiliation{\mbox{ Department of Engineering Physics, Tsinghua University, Beijing, China 100084}}
\affiliation{\mbox{ Key Laboratory of Particle and Radiation Imaging (Tsinghua University), Ministry of Education, Beijing, China 100084}}
\affiliation{\mbox{ College of Physics, Sichuan University, Chengdu, China 610065}}

\author{Zhiyuan Wang}
\affiliation{\mbox{ Department of Engineering Physics, Tsinghua University, Beijing, China 100084}}
\affiliation{\mbox{ Key Laboratory of Particle and Radiation Imaging (Tsinghua University), Ministry of Education, Beijing, China 100084}}

\author{Xinlong Cao}
\affiliation{\mbox{ Department of Engineering Physics, Tsinghua University, Beijing, China 100084}}
\affiliation{\mbox{ Key Laboratory of Particle and Radiation Imaging (Tsinghua University), Ministry of Education, Beijing, China 100084}}

\author{Zhuoxuan Liu}
\affiliation{\mbox{ Department of Engineering Physics, Tsinghua University, Beijing, China 100084}}
\affiliation{\mbox{ Key Laboratory of Particle and Radiation Imaging (Tsinghua University), Ministry of Education, Beijing, China 100084}}

\author{Lianmin Zheng}
\affiliation{\mbox{ Department of Engineering Physics, Tsinghua University, Beijing, China 100084}}
\affiliation{\mbox{ Key Laboratory of Particle and Radiation Imaging (Tsinghua University), Ministry of Education, Beijing, China 100084}}

\author{Yingchao Du}
\affiliation{\mbox{ Department of Engineering Physics, Tsinghua University, Beijing, China 100084}}
\affiliation{\mbox{ Key Laboratory of Particle and Radiation Imaging (Tsinghua University), Ministry of Education, Beijing, China 100084}}

\author{Wenhui Huang}
\affiliation{\mbox{ Department of Engineering Physics, Tsinghua University, Beijing, China 100084}}
\affiliation{\mbox{ Key Laboratory of Particle and Radiation Imaging (Tsinghua University), Ministry of Education, Beijing, China 100084}}

\author{Chuanxiang Tang}
\affiliation{\mbox{ Department of Engineering Physics, Tsinghua University, Beijing, China 100084}}
\affiliation{\mbox{ Key Laboratory of Particle and Radiation Imaging (Tsinghua University), Ministry of Education, Beijing, China 100084}}

\author{Renkai Li}
\altaffiliation{Email: lirk@tsinghua.edu.cn}
\affiliation{\mbox{ Department of Engineering Physics, Tsinghua University, Beijing, China 100084}}
\affiliation{\mbox{ Key Laboratory of Particle and Radiation Imaging (Tsinghua University), Ministry of Education, Beijing, China 100084}}

\date{\today}

\begin{abstract} 


Pushing the brightness limit of electron sources requires simultaneously minimizing the intrinsic emittance and maximizing the accelerating field. Alkali-antimonide photocathodes exhibit excellent properties at low fields, yet their high-field photoemission physics remains poorly understood owing to acute vacuum sensitivity. Here, we demonstrate robust photoemission from alkali-antimonide photocathodes in a radio-frequency gun at peak fields exceeding 100~MV/m, with quantum efficiency maintained above 1\% for more than two weeks, enabling the first systematic measurements of their high-field brightness limits. The measured field dependence of the intrinsic emittance provides direct insight into photoemission physics at high fields. Near-threshold photoemission further reduces the intrinsic emittance, increasing the attainable brightness, and reveals constraints imposed by surface roughness. This work opens the high-field regime for advanced photocathodes, extending the brightness limits of electron sources.


\end{abstract}

\maketitle

The pursuit of higher electron-beam brightness has long been a central thrust of beam physics and accelerator technology. Advances in beam brightness have enabled a broad range of modern scientific instruments and experimental methods, including short-wavelength free-electron lasers, inverse Compton scattering sources, ultrafast electron scattering, and electron cooling for colliders~\cite{dowell2010cathode,musumeci2018advances,schaber2023review}. Their capabilities and future advances critically depend on the achievable beam brightness, which is fundamentally governed by the electron source~\cite{Bazarov2009Maximum, filippetto2014maximum, wang2016report}. The brightness limit set by beam generation scales as 
\begin{eqnarray}\label{scale}
  B_{m}=\frac{Q}{\epsilon^2_n}\propto\frac{E^{s}}{\textrm{MTE}},
\end{eqnarray}
where $Q$ is the bunch charge, $\epsilon_n$ is the normalized emittance, $E$ is the emission electric field strength, and MTE denotes the mean transverse energy of photoelectrons. The exponent $s$ is controlled by the initial aspect ratio of the electron bunch, ranging from 1 for pancake beams and $3/2$ for cigar beams~\cite{Bazarov2009Maximum, filippetto2014maximum}. Maximizing the brightness limit therefore requires the combination of low MTE and high emission and accelerating fields in the electron source. Quantum efficiency (QE) is another key figure-of-merit for photocathodes, enabling the high beam current and average brightness required to meet the signal-to-noise ratio and beam-power demands of modern accelerator applications. The ability to operate under visible-light illumination, which substantially relaxes the laser shaping and power requirements, together with long operational lifetime, are additional important attributes of photocathodes for high-performance and reliable operation.

Alkali antimonides are among the most extensively studied class of photocathodes and have demonstrated outstanding performance at low accelerating fields. QE of $\sim5\%$ under 532-nm laser illumination and intrinsic emittance as low as 0.4–0.5 mm mrad/mm have been reported at fields below 10 MV/m~\cite{vecchione2011low, bazarov2011thermal}. These characteristics make them leading candidates for further enhancing electron source brightness. Substantial work at low field has characterized the wavelength dependence of QE and intrinsic emittance~\cite{vecchione2011low, bazarov2011thermal, mohanty2023development}. However, the photoemission physics of these materials at significantly higher emission fields is largely unexplored, and whether their favorable low-field properties are retained at high fields has yet to be verified.

\begin{figure*}
    \centering
    \includegraphics[width=1\linewidth]{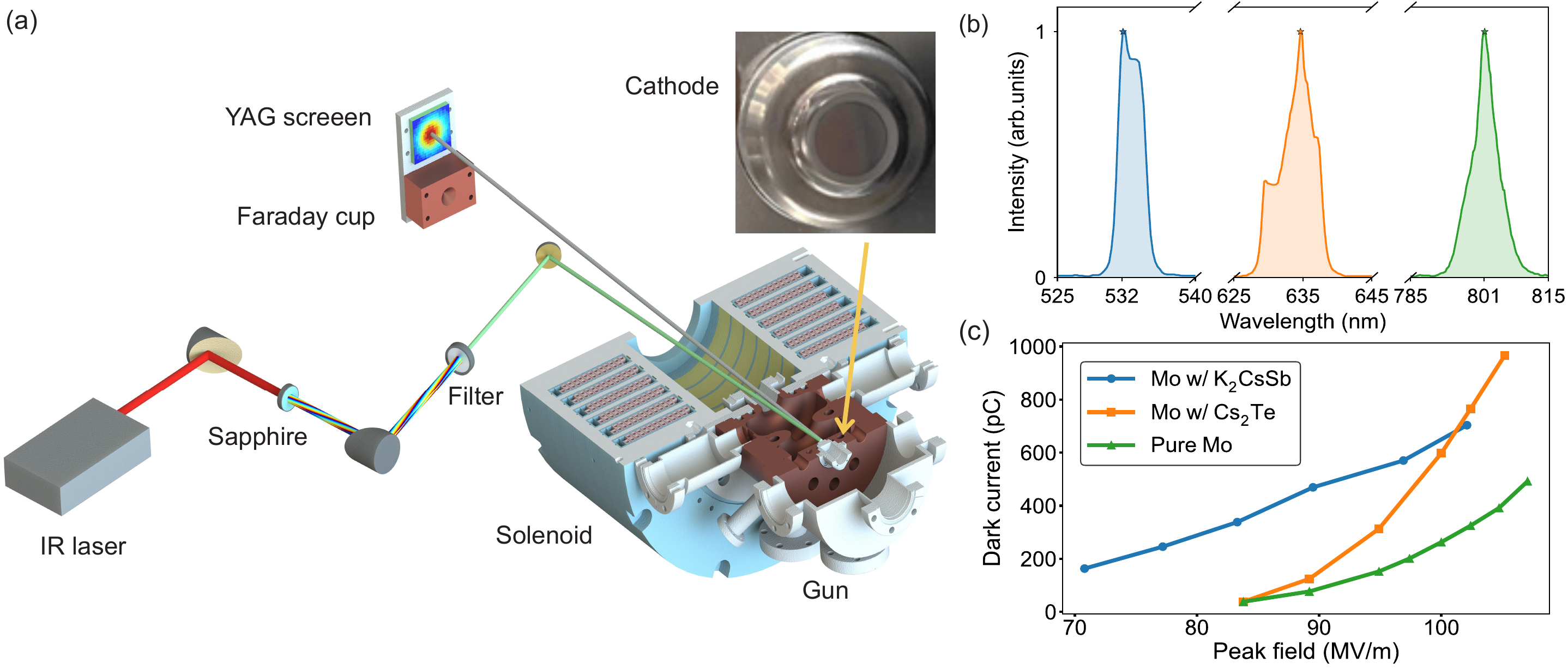}
    \caption{Schematic of the experiment. (a) Layout of the measurement apparatus. (b) Measured spectra of 532-nm, 635-nm, and 800-nm laser. (c) Measured dark current as a function of peak electric field for K$_2$CsSb, Cs$ _2$Te, and Mo plug.}
    \label{beamline}
    \end{figure*}

Practical high-field electron guns present a challenging vacuum environment owing to enhanced field-induced outgassing and limited conductance in compact geometries. Alkali-antimonide photocathodes are particularly sensitive to residual gases, making their operational lifetime strongly dependent on vacuum quality~\cite{kuriki2017lifetime, panuganti2021synthesis, kuriki2016study, Ding2017Temperature}. Recent explorations in S-band and L-band guns~\cite{martinicharacterization, garciaipac2024wepc67, mohantyipac2025} have reported cathode lifetimes limited to a few tens of hours, and their susceptibility to vacuum degradation remains a major constraint on reliable characterization and sustained operation in high-field sources. Intrinsic emittance, which together with the emission field sets the beam-brightness limit, may increase at high field due to Schottky barrier lowering and transverse fields associated with surface roughness~\cite{Qian2012Experimental, gevorkyan2018effects, zhang2015roughness,feng2017near}. The field dependence of the intrinsic emittance, and hence the attainable brightness, of alkali-antimonide photocathodes at gradients of several tens of MV/m or higher remains to be experimentally established.

In this Letter, we demonstrate that the brightness limit can be extended by operating a K$_2$CsSb photocathode in an ultrahigh‑vacuum S‑band RF gun with peak fields exceeding 100 MV/m. The cathode exhibited QE above 1$\%$ under 532-nm laser illumination for a record lifetime exceeding two weeks. Using a transverse momentum imaging method, we characterized the intrinsic emittance as a function of the emission field and observed an increase in the brightness limit. Furthermore, we explored near-threshold photoemission using a 635-nm laser, achieving lower intrinsic emittance than in the 532-nm case, thereby further improving the brightness limit and demonstrating a clear advantage over previously reported photocathodes. White‑light interferometry was carried out to analyze the cathode surface morphology after high‑field operation, revealing surface roughness features consistent with those inferred from the measured field dependence of the intrinsic emittance.

The simultaneous realization of ultrahigh vacuum and high accelerating field is essential for the stable operation of alkali-antimonide photocathodes. To overcome this challenge, an S-band electron gun with enhanced vacuum conductance was developed, incorporating a modified cathode plate with tailored perforations and a dedicated upstream pumping cavity~\cite{huang2023ultrahigh}. The system reached a base pressure of $7.5 \times 10^{-9}$ Pa and sustained operation above 100 MV/m while maintaining a pressure of $1.7 \times 10^{-8}$ Pa. Numerical modeling was performed to evaluate the pressure at the cathode surface, which is found to be approximately $3.0 \times 10^{-8}$ Pa. These conditions provide a platform for quantitative studies of cathode performance under stable high-field operation.

The experiments were carried out at the Tsinghua Thomson Scattering X-ray Source facility~\cite{tang2009tsinghua}, with the schematic shown in Fig.~\ref{beamline}. Cathode drive lasers at 532 nm and 635 nm were obtained via supercontinuum generation in sapphire from an 800-nm drive pulse~\cite{imran2012intensity, dubietis2017ultrafast} and selected using narrow band-pass filters. The measured spectra are shown in Fig.~\ref{beamline}(b). Electron beams were generated by near-normal-incidence laser illumination of the cathode and accelerated to multi-MeV energies by the ultrahigh-vacuum S-band gun. A solenoid immediately downstream of the gun was used for beam focusing and momentum-imaging emittance measurements. A movable Faraday cup and a YAG screen, located at 134 cm from the cathode, were used for bunch charge measurement and transverse profile characterization. K$_2$CsSb cathodes were synthesized via sequential deposition~\cite{mohanty2023development, schubert2016bi} and transferred into the gun using a vacuum suitcase and load-lock system. 

K$_2$CsSb photocathodes have a relatively low photoemission threshold of $\sim$2 eV. This property, while substantially relaxing laser shaping and power requirements, increases their susceptibility to field emission, leading to higher dark current and posing a challenge for high-field operation. Fig.~\ref{beamline}(c) shows the measured dark current as a function of gun peak field for K$_2$CsSb and Cs$_2$Te cathodes, together with a pure Mo plug, obtained with an RF pulse duration of 1.6 $\mu$s. The dark current from the K$_2$CsSb cathode is consistently higher than that from the pure Mo plug, but exhibits weaker field dependence than the commonly used Cs$_2$Te, which may be attributed to a lower field enhancement factor of the K$_2$CsSb cathode. Despite these characteristics, stable operation at 100 MV/m was maintained for the K$_2$CsSb cathode, with the dark current remaining compatible with high-field operation.

\begin{figure}[!tbp]
   \includegraphics*[width=1\columnwidth]{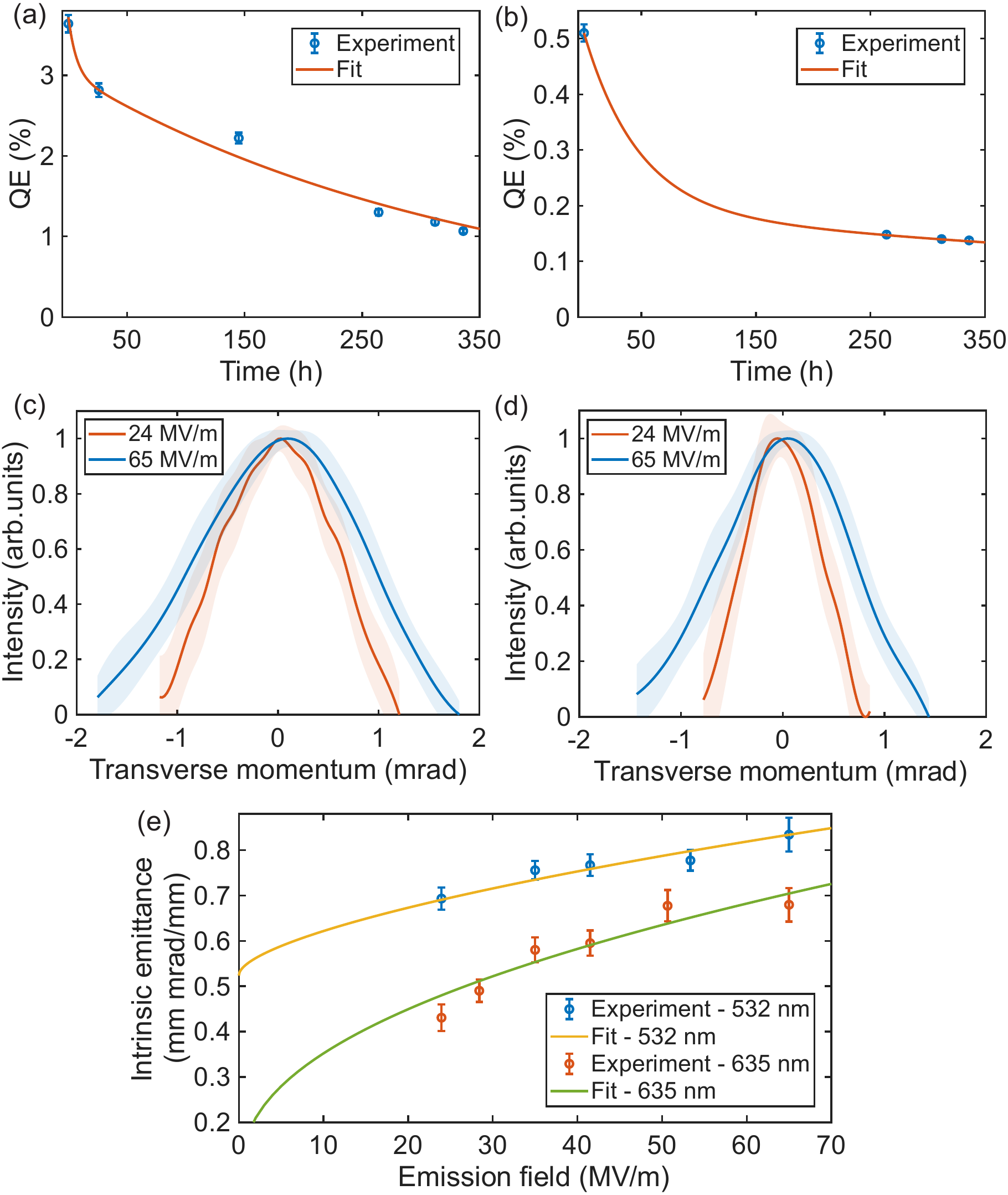}
   \caption{QE evolution with (a) 532-nm and (b) 635-nm laser illumination. Measured transverse momentum distribution of photoelectrons with (c) 532-nm and (d) 635-nm laser illumination at a field strength of 24 MV/m and 65 MV/m, respectively. The shaded region represents the error bars from measurements. (e) Measured intrinsic emittance against the emission electric field with 532-nm and 635-nm lasers.}\label{e532nm}
\end{figure}

K$_2$CsSb photocathodes exhibit a high QE, which depends on the difference between the drive-laser photon energy and the electron affinity. The electron affinity can increase due to surface degradation under the complex vacuum conditions encountered during high-power RF conditioning, leading to a reduction in QE. After cathode insertion into the gun, conditioning to fields of 100 MV/m was completed within $\sim$4 hours. A QE of 3.6$\%$ was then measured under 532-nm laser illumination, as shown in Fig.~\ref{e532nm}(a), compared to an initial QE of 5.0$\%$ measured in the cathode preparation chamber. This indicates that the QE is well preserved during cathode transfer and subsequent RF conditioning.

Achieving extended operational lifetime of alkali-antimonide photocathodes remains a key challenge for exploiting their full potential. The substantially improved vacuum conditions in the present gun appear to mitigate electron-affinity increase and surface passivation, thereby enabling sustained high-field operation. Experimental results show that the photocathode maintains a QE exceeding 1$\%$ for more than two weeks under 532-nm illumination at an emission field of 50 MV/m, i.e. the typical 30$^\circ$ launch RF phase with a peak field of 100 MV/m, as shown in Fig.~\ref{e532nm}(a). The measurements were performed with several pC bunch charges and a repetition rate of 10 Hz, corresponding to a low average current, for which beam-induced effects on QE evolution are negligible. The QE degradation slows significantly as it approaches the 1$\%$ level. This establishes a regime of combined accelerating field and lifetime not previously demonstrated for alkali antimonides~\cite{Ding2017Temperature,mohantyipac2025, garciaipac2024wepc67, martinicharacterization}, providing a viable pathway toward stable operation of this class of photocathodes in high-field RF guns.


This capability enables a quantitative investigation of whether the low MTE of K$_2$CsSb observed at low fields is preserved at high fields. The intrinsic emittance $\epsilon_i$ can be decomposed into a zero-field component and field-dependent contributions. The zero-field term scales with the difference between the drive photon energy $\hbar\omega$ and the emission threshold $\phi$. An applied field $E$ modifies $\epsilon_i$ via the Schottky effect and surface roughness: the former lowers the vacuum barrier, increasing the excess energy of emitted electrons that scales with $\sqrt{E}$, while the latter induces transverse fields proportional to $E$ and dominates at high fields. The intrinsic emittance can thus be written as~\cite{feng2017near}:
\begin{eqnarray}\label{growth}
  \epsilon_i &=& \sqrt{\dfrac{1}{3m_ec^2}(\hbar\omega-\phi+e\sqrt{\dfrac{eE}{4\pi\varepsilon_0}}+\dfrac{3e\pi^2a^2}{2\lambda}E)},
\end{eqnarray}
where $m_ec^2$ is the electron rest energy, $e$ is the elementary charge, $\varepsilon_0$ is the vacuum permittivity, and $a$ and $\lambda$ are the surface roughness amplitude and spatial period, respectively.

The intrinsic emittance of the K$_2$CsSb photocathode under 532-nm laser illumination was characterized using the transverse momentum imaging method~\cite{PhysRevAccelBeams.23.043401}. Under optimal solenoid focusing, contributions from the initial transverse profile at the cathode are effectively eliminated, and the image recorded on the YAG screen directly reflects the transverse momentum distribution. As shown in Fig.~\ref{e532nm}(c), the momentum profiles at different accelerating fields indicate that, as the field increases, a larger fraction of electrons exhibit higher transverse momenta, leading to a broadening of the distribution.

The measured intrinsic emittance exhibits a clear field dependence, as shown in Fig.~\ref{e532nm}(e), increasing from 0.69 mm mrad/mm at 24 MV/m to 0.83 mm mrad/mm at 65 MV/m. The field at photoemission was tuned by adjusting the RF phase and the peak field. The approximately Gaussian profiles at all fields suggest that the field-induced contribution acts as an additional Gaussian broadening mechanism in transverse momentum. The experimental data are well described by Eq.~(\ref{growth}), from which a surface roughness coefficient $\pi^2 a^2 / 2\lambda = 1.8 \pm 0.3$ nm is extracted. The fitted zero-field intrinsic emittance is 0.52 mm mrad/mm, in agreement with previous low-field measurements~\cite{bazarov2011thermal}, corresponding to an MTE of 138 meV.

Operating the drive laser near the photoemission threshold is a key strategy for approaching the intrinsic emittance limit of photocathodes. Reducing the excess energy of emitted electrons suppresses intrinsic emittance, albeit at the expense of lower QE and enhanced sensitivity to the accelerating field. To experimentally quantify this trade-off, we investigate photoemission from the K$_2$CsSb photocathode under 635-nm (1.95 eV) laser illumination, characterizing its behavior in the near-threshold regime.


The initial QE of the K$_2$CsSb photocathode under 635-nm laser illumination was $5.1 \times 10^{-3}$ at an emission field of 50 MV/m, as shown in Fig.\;\ref{e532nm}(b). In close proximity to the emission threshold, the QE exhibits enhanced sensitivity to threshold variations. After two weeks of operation, the QE decreased to $1.3\times10^{-3}$, with a diminishing degradation rate, indicating saturation of the threshold shift~\cite{Huang2019photoemission}. These results confirm the feasibility of stable long-term operation of K$_2$CsSb in the near-threshold regime at QE levels on the order of $10^{-3}$, sufficient for a broad range of high-brightness beam applications, including UED and XFELs.


Near-threshold photoemission effectively reduces the intrinsic emittance,  which was measured to be 0.43 mm mrad/mm at 24 MV/m, as shown in Fig.\;\ref{e532nm}(c), representing a $\sim$40\% reduction compared to that with 532-nm illumination. At a higher emission field of 65 MV/m, the intrinsic emittance increases to 0.68 mm mrad/mm, consistent with the scaling predicted by Eq.\;(\ref{growth}). Fitting yields a surface roughness coefficient of $\pi^2 a^2 / 2\lambda = 2.4 \pm 0.5$ nm, while the zero-field intrinsic emittance is approximately 0.2 mm mrad/mm, consistent with the thermal limit at room temperature.

\begin{figure}[!tbp]
   \includegraphics*[width=0.8\columnwidth]{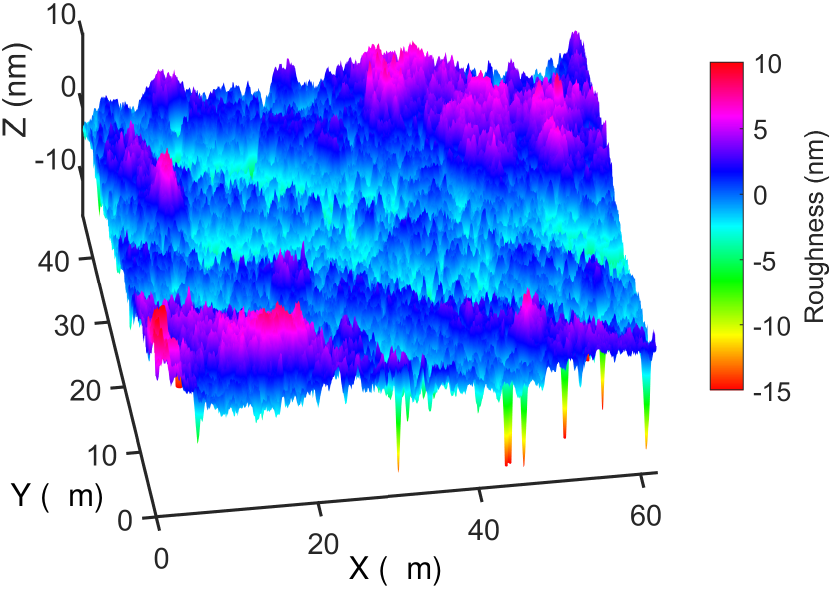}
   \caption{Surface morphology of the K$_2$CsSb photocathode after high-field operation, measured by white-light interferometry. }\label{roughness}
\end{figure}

The measured field dependence of the intrinsic emittance at 532 nm and 635 nm indicates that surface roughness is a significant factor in the high-field performance of K$_2$CsSb photocathodes, with direct implications for brightness limits. The roughness arises from two primary sources: the deposition methodology and potential damage incurred during high-field operation. The sequential deposition technique employed in this study is known to produce higher surface roughness in K$_2$CsSb compared to co-deposition methods~\cite{feng2017near}. During operation at 100 MV/m—a previously unexplored regime—breakdown events can occur, triggering ion bombardment cascades that generate localized surface features leading to larger roughness.


To quantify its surface morphology, the K$_2$CsSb photocathode was characterized using white-light interferometry following high-field operation. The cathode was briefly exposed to ambient atmosphere during transfer from the vacuum suitcase to the interferometry station. The analyzed region is located at the center of the cathode plug, coinciding with the laser-illuminated area during operation. As shown in Fig.\;\ref{roughness}, the operated film exhibits an rms surface roughness of 20.6 nm, with an average peak spacing of approximately 1.0 $\mu$m. The corresponding roughness scaling coefficient, $\pi^2 a^2 / 2\lambda$, is determined to be 2.1 nm, consistent with values extracted from intrinsic emittance fits under both 532-nm ($1.8 \pm 0.3$ nm) and 635-nm ($2.4 \pm 0.5$ nm) illumination. This observation suggests that, to fully leverage the low intrinsic emittance of K$_2$CsSb under high fields, co-deposition with near-atomic-scale control is desirable to suppress the stochastic formation of pronounced surface roughness associated with sequential deposition methods. 

\begin{figure}[!tbp]
   \hspace*{-0.3cm}\includegraphics*[width=1.02\columnwidth]{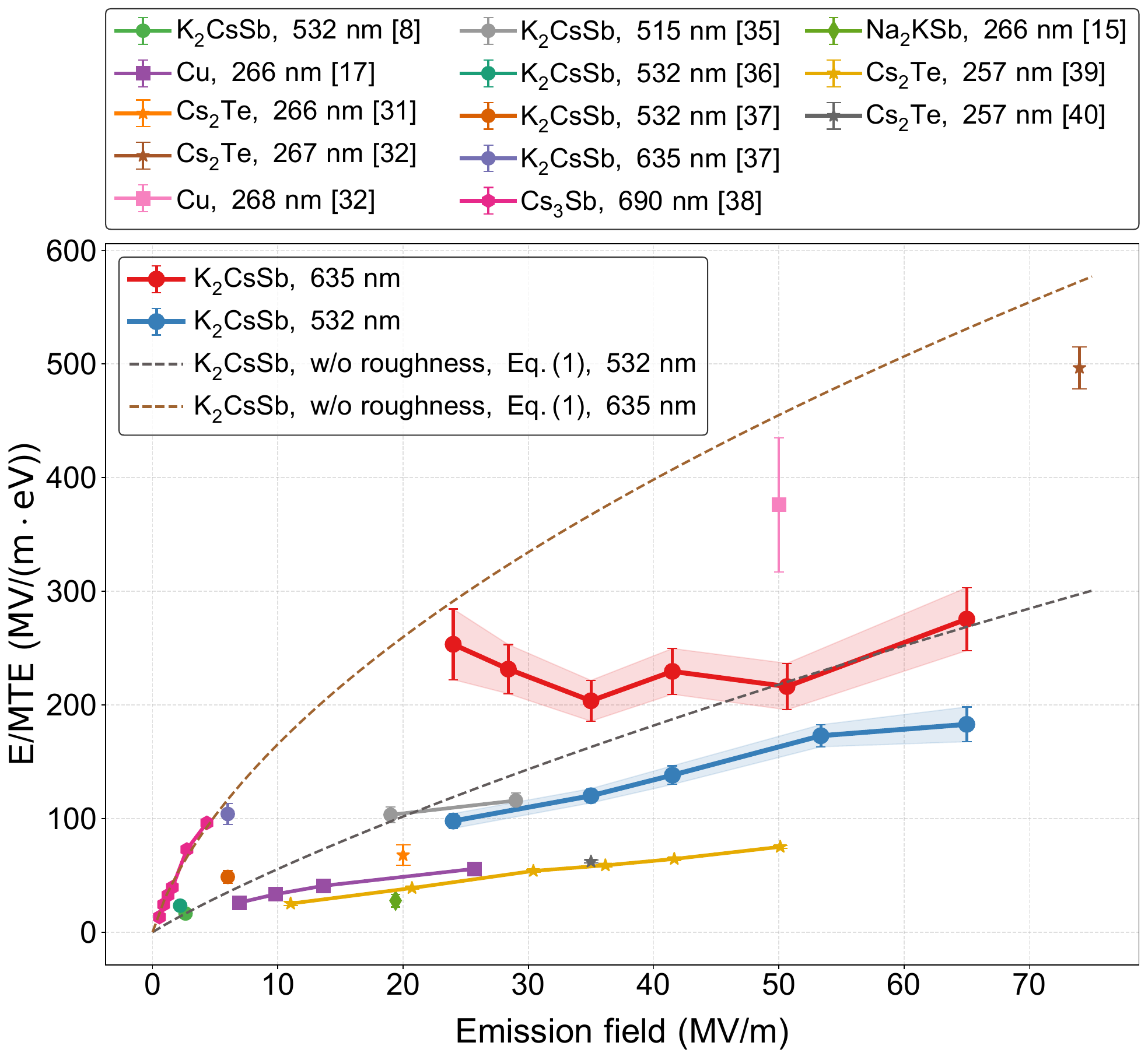}
   \caption{Comparison of reported brightness limits for several photocathode sources, with cathode materials and drive-laser wavelengths indicated. Entries in the legend marked separately denote the results of this work.~\cite{mohanty2022development, jlab2020thermal, PKU2024thermal, cultrera2015cold, bazarov2011thermal, garciaipac2024wepc67, huang2019test, zheng2018overestimation, filippetto2015cesium, prat2015measurements, Qian2012Experimental}.}\label{brightnesslimit}
\end{figure}

We evaluate the brightness limit of K$_2$CsSb from the measured intrinsic emittance at different emission fields and compare it with recently reported values for a range of photocathode sources, including alkali antimonides in DC and RF injectors, as well as Cs$_2$Te and Cu in RF injectors, as shown in Fig.~\ref{brightnesslimit}. All data points are calculated using Eq.~(\ref{scale}) with exponent $s=1$. The comparison shows that K$_2$CsSb achieves a notably higher brightness limit than the widely used high-field-compatible photocathodes Cs$_2$Te and Cu. Most reported measurements for Cs$_2$Te and Cu yield brightness limits below 100 MV/(m$\cdot$eV)~\cite{PhysRevAccelBeams.23.043401, PhysRevAccelBeams.25.053401, zheng2020rapid, Xiang1302027, filippetto2015cesium, prat2015measurements}. Cs$_2$Te typically exhibits stable percent-level QE, whereas Cu operates at QE levels of $10^{-4}$ or below; both are driven by UV lasers. Two data points reported in Ref.~\cite{prat2015measurements} are notable exceptions and lie well above the values reported for similar cathode materials in other sources. For alkali antimonides in DC and SRF injectors, the lower intrinsic emittance is offset by the lower emission field, which limits the attainable brightness. By contrast, K$_2$CsSb combines low intrinsic emittance with high-field operation, yielding a brightness limit of up to 183 MV/(m$\cdot$eV) under 532-nm laser illumination. Near-threshold illumination at 635 nm further increases the brightness limit to 276 MV/(m$\cdot$eV). Dashed lines in Fig.~\ref{brightnesslimit} indicate that the brightness limit could be further enhanced if cathode surface roughness were reduced to a negligible level, reaching approximately 500 MV/(m$\cdot$eV) at the demonstrated emission field, as predicted by Eq.~(\ref{scale}). Achieving near-atomically smooth surfaces through epitaxial growth will be essential to suppress the roughness-induced emittance growth at high gradients~\cite{parzyck2022single} and unlock the ultralow subthermal emittance predicted by the electron refraction model~\cite{rozhkov2026subthermal}.


In summary, this work reports a new regime for photocathode electron sources, in which alkali-antimonide semiconductor cathodes under RF accelerating fields up to 100 MV/m demonstrate enhanced brightness limits. We experimentally studied the key characteristics of K$_2$CsSb photocathodes under high field, including QE and intrinsic emittance, in a newly developed RF gun with ultrahigh vacuum performance. The cathode sustained operation for more than two weeks while maintaining a QE above 1$\%$ under 532-nm laser illumination, demonstrating its compatibility with high accelerating fields. Measurements of the intrinsic emittance reveal a higher accessible brightness limit than that of conventional photocathodes. We further explored near-threshold photoemission at 635 nm, where reduced intrinsic emittance and a sustainable QE above $10^{-3}$ together yield a more than twofold increase in the brightness limit relative to reported values. The observed field-dependent emittance growth, correlated with white-light interferometry measurements, identifies surface roughness as the dominant microscopic mechanism currently limiting further brightness enhancement. These results establish alkali-antimonide photocathodes under high field as a new route toward high-brightness electron sources and provide an ultrahigh-vacuum, high-gradient platform for studying exotic and engineered quantum-material photocathodes in practical electron source conditions.

\begin{acknowledgments}
The work was partially supported by the National Key Research and Development Program of China (Grant No. 2022YFA1603400 and 2022YFA1602200) and the National Natural Science Foundation of China (Grant No. 12505175).
\end{acknowledgments}

\bibliography{main}

\end{document}